\documentclass[aps,pra,amsmath,twocolumn,amssymb,superscriptaddress]{revtex4-2}

\usepackage{amssymb,amsfonts,amsmath}
\usepackage{color}
\usepackage[apple mac]{inputenc}
\usepackage[english]{babel}
\usepackage{times}
\usepackage{latexsym}
\usepackage{fancyhdr}
\usepackage{verbatim}
\usepackage{graphicx}
\usepackage{cancel}
\usepackage{epsf}
\usepackage{subfigure}
\usepackage{bm}
\usepackage{epstopdf}

\usepackage[colorlinks=true,
            linkcolor=magenta,
            urlcolor=blue,
            citecolor=blue]{hyperref}

\definecolor{Nathanblue}{rgb}{0.,0.24,0.51}
\newcommand{\blue}{\color{Nathanblue}}

\definecolor{orange}{rgb}{1.,0.2,0.2}

\usepackage{mathtools}

\def\be{\begin{equation}}
\def\ee{\end{equation}}
\def\bs#1{\boldsymbol{#1}}

\begin{document}

\title{{\blue Relating the Hall conductivity to the many-body Chern number \\ using Fermi's Golden rule and Kramers-Kronig relations}}

\author{Nathan Goldman}
\email[]{nathan.goldman@lkb.ens.fr}
\affiliation{Laboratoire Kastler Brossel, Coll\`ege de France, CNRS, ENS-PSL University, Sorbonne Universit\'e, 11 Place Marcelin Berthelot, 75005 Paris, France
}
\affiliation{CENOLI,
Universit\'e Libre de Bruxelles, CP 231, Campus Plaine, B-1050 Brussels, Belgium}
\author{Tomoki Ozawa}
\email[]{tomoki.ozawa.d8@tohoku.ac.jp}
\affiliation{Advanced Institute for Materials Research (WPI-AIMR), Tohoku University, Sendai 980-8577, Japan }

\begin{abstract}

This pedagogical piece provides a surprisingly simple demonstration that the quantized Hall conductivity of correlated insulators is given by the many-body Chern number, a topological invariant defined in the space of twisted boundary conditions. In contrast to conventional proofs, generally based on the Kubo formula, our approach entirely relies on combining Kramers-Kronig relations and Fermi's golden rule within a circular-dichroism framework. This pedagogical derivation illustrates how the Hall conductivity of correlated insulators can be determined by monitoring single-particle excitations upon a circular drive, a conceptually simple picture with direct implications for quantum-engineered systems, where excitation rates can be directly monitored.

\end{abstract}

\date{\today}

\maketitle

\section{Introduction}

The integer and fractional quantum Hall effects were discovered in the early 1980's~\cite{Klitzing:1980,Tsui:1982}, by investigating the Hall conductivity of two-dimensional electron gases subjected to a strong magnetic field. The quantum Hall effects refer to the remarkable quantization of the Hall conductivity observed in these settings:~$\sigma_H/\sigma_0\!=\!\nu$, where $\nu$ is an integer or a fractional number, and where $\sigma_0\!=\!e^2/h$ is the quantum of conductance. Soon after these discoveries, it became clear that the quantization of the Hall conductivity was deeply rooted in the topological nature of the underlying many-body quantum state~\cite{thouless1982quantized,simon1983holonomy,avron1983homotopy}. Since then, the study of topological quantum matter has become an important subject in condensed matter physics~\cite{KaneMele1,KaneMele2,BernevigHughesZhang,BernevigZhang,Konig:2007}, but also in other fields, such as photonics~\cite{ozawa2019topological} and ultracold quantum gases~\cite{cooper2019topological}. 

Both integer and fractional quantum Hall states can be characterized by a topological invariant, known as the many-body Chern number~\cite{niu1985quantized,tao1986impurity,avron1985quantization}. This topological invariant is constructed from the many-body ground state, as defined over an abstract parameter space associated with generalized (twisted) boundary conditions. Generally, the relation between the quantized Hall conductivity and the many-body Chern number is obtained using linear-response theory, through the Kubo formula~\cite{niu1985quantized,tao1986impurity,avron1985quantization}. Although the derivation based on Kubo's formula is well established, the importance of this topological phenomenon may warrant an alternative look from a different angle.

For the integer quantum Hall effect, the relation between the Hall conductivity and the Chern number can be derived through the Streda formula~\cite{kohmoto1989zero}. However, a generalization of this Streda approach to correlated insulators, such as fractional quantum Hall states, remains elusive. We note that the Hall conductivity of correlated insulators can be related to another many-body invariant, known as the Ishikawa-Matsuyama invariant~\cite{ishikawa1986magnetic,ishikawa1987microscopic,imai1990field,wang2010topological,gurarie2011single,essin2011bulk,wang2012simplified}, using the Streda formula~\cite{gavensky2023connecting}. However, to the best of our knowledge, a formal relation between the Ishikawa-Matsuyama invariant and the many-body Chern number of fractional quantum Hall states is lacking.

In this pedagogical piece, we provide a surprisingly simple demonstration of the relation between the quantized Hall conductivity and the many-body Chern number of correlated insulators, which entirely relies on combining Kramers-Kronig relations and Fermi's golden rule. Kramers-Kronig relations relate a DC (Hall) conductivity to an optical (dissipative) response. In the context of circular dichroism, this relation allows one to express the Hall conductivity in terms of excitation rates upon a circular drive defined in the space of twisted boundary conditions (or flux space). This alternative approach, which is described in detail in the following Sections, is inspired from a series of theoretical and experimental works on topological circular-dichroic responses~\cite{tran2017probing,schuler2017tracing,tran2018quantized,asteria2019measuring,ozawa2019probing,repellin2019detecting,pozo2019quantization,tan2019experimental,yu2020experimental,midtgaard2020detecting,schuler2020circular,schuler2020local,klein2021interacting,le2022global,chen2022synthetic,yu2022experimental}.

The organization of this work is as follows. In Sec.~\ref{sec:mbcn}, we define the many-body Chern number in a general setting. In Sec.~\ref{sec:mbcntwist}, we define the many-body Chern number in twist-angle (or flux) space. Readers who are familiar with these notions can skip Secs.~\ref{sec:mbcn} and \ref{sec:mbcntwist}, and directly go to Sec.~\ref{sec:cd}, where we relate the many-body Chern number to a circular-dichroic response using Fermi's golden rule. In Sec.~\ref{sec:kk}, using the Kramers-Kronig relation, we relate the circular-dichroic response to the Hall conductivity, thus establishing the connection between the many-body Chern number and the Hall conductivity. We provide concluding remarks in Sec.~\ref{sec:conclusion}.

\section{The many-body Chern number}
\label{sec:mbcn}

\begin{figure}[h!]
{\scalebox{0.22}
{\includegraphics{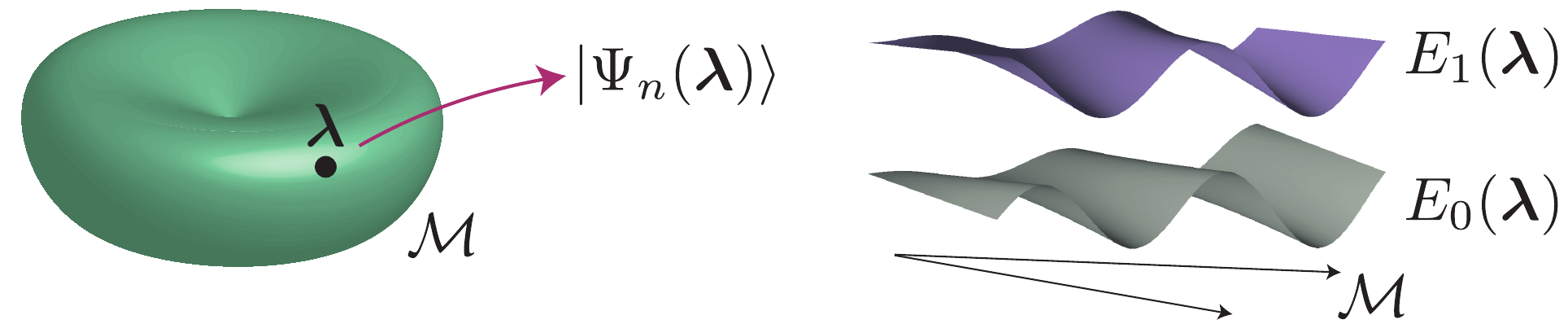}}}
\caption{Sketch of the general setting, illustrating eigenvectors $\vert \Psi_n (\bs{\lambda}) \rangle$ and energies $E_n (\boldsymbol\lambda)$ over a 2D parameter space $\mathcal{M}$.}
\label{fig:illustration}
\end{figure}

We first consider a generic system described by a Hamiltonian $\hat H (\bs{\lambda})$ over a two-dimensional parameter space, $\bs{\lambda}=(\lambda^1,\lambda^2) \in \mathcal{M}$. The local eigenstates are defined from the eigenvalue problem: 
\begin{align}
\hat H (\bs{\lambda}) \vert \Psi_n (\bs{\lambda}) \rangle = E_n (\bs{\lambda}) \vert \Psi_n (\bs{\lambda}) \rangle. \label{eq:firstequation}
\end{align}
The energy eigenvalues $E_n (\boldsymbol\lambda)$ form the band structure, whereas the $\boldsymbol\lambda$-dependence of the eigenvector $\vert \Psi_n (\bs{\lambda}) \rangle$ determines the geometric properties of the band.
This general setting is schematically illustrated in Fig.~\ref{fig:illustration}.

If the ground-state band $E_0(\bs{\lambda})$ is well isolated (non-degenerate case) and $\mathcal{M}$ is closed, then one can define the Chern number of the ground-state band as
\begin{align}
&\nu_{\rm{Ch}} = \frac{1}{2 \pi} \int_{\mathcal{M}} \mathcal{F}_{12} (\bs{\lambda}) \, \rm{d}^2\lambda \in \mathbb{Z} , \notag \\
\end{align}
where
\begin{align}
&\mathcal{F}_{12} (\bs{\lambda})= i  \Biggl ( \, \bigg \langle \frac{\partial \Psi_0}{\partial \lambda^{1}} \bigg \vert \frac{\partial \Psi_0}{\partial \lambda^{2}} \bigg \rangle - \bigg \langle \frac{\partial \Psi_0}{\partial \lambda^{2}} \bigg \vert \frac{\partial \Psi_0}{\partial \lambda^{1}} \bigg \rangle  \Biggr ) , \label{eq:berrycurvature}
\end{align}
is the Berry curvature. Here we assume that the state is normalized, $\langle \Psi_0 | \Psi_0 \rangle = 1$.

For later convenience, we now derive an alternative expression for the Berry curvature, which does not involve derivatives of the states. Taking the derivative of Eq.~\eqref{eq:firstequation} for the ground state with respect to a parameter $\lambda^i$, we obtain
\begin{align}
	\frac{\partial \hat{H}}{\partial \lambda^i}|\Psi_0\rangle + \hat{H}\left| \frac{\partial \Psi_0}{\partial \lambda^i}\right\rangle = \frac{\partial E_0}{\partial \lambda^i}|\Psi_0\rangle + E_0 \left| \frac{\partial \Psi_0}{\partial \lambda^i}\right\rangle.
\end{align}
Acting with a state $\langle \Psi_n|$, which is different from the ground state, from the left, we obtain
\begin{align}
	\langle \Psi_n | \frac{\partial \hat{H}}{\partial \lambda^i}|\Psi_0\rangle  = (E_0 - E_n)\langle \Psi_n \left| \frac{\partial \Psi_0}{\partial \lambda^i}\right\rangle.
\end{align}
Using the identity $1 = |\Psi_0 \rangle \langle \Psi_0 | + \sum_{n\neq 0} |\Psi_n\rangle \langle \Psi_n |$, the Berry curvature \eqref{eq:berrycurvature} can be rewritten as
\begin{align}
	\mathcal{F}_{12}(\boldsymbol\lambda)
	=
	i \sum_{n\neq 0}& \left( \, \left \langle \frac{\partial \Psi_0}{\partial \lambda^{1}} \right|\Psi_n\rangle\langle\Psi_n \left| \frac{\partial \Psi_0}{\partial \lambda^{2}} \right\rangle
	\right.
	\notag \\
	&
	\left.-
	\left\langle \frac{\partial \Psi_0}{\partial \lambda^{2}} \right|\Psi_n\rangle\langle\Psi_n \left|\frac{\partial \Psi_0}{\partial \lambda^{1}} \right\rangle  \right) 
	\notag \\
	=
	i \sum_{n\neq 0}& \frac{1}{(E_n - E_0)^2}\left( \, \langle \Psi_0 | \frac{\partial \hat{H}}{\partial \lambda^1} | \Psi_n\rangle\langle\Psi_n | \frac{\partial \hat{H}}{\partial \lambda^2}| \Psi_0\rangle
	\right.
	\notag \\
	&
	\left.-
	\langle \Psi_0 | \frac{\partial \hat{H}}{\partial \lambda^2} | \Psi_n\rangle\langle\Psi_n | \frac{\partial \hat{H}}{\partial \lambda^1}| \Psi_0\rangle  \right). \label{eq:berrycurvature2}
\end{align}
The final expression involves the derivative of the Hamiltonian with respect to the parameters, but not the derivatives of the states. We will use this expression in the next Sections to connect the circular-dichroic response and the many-body Berry curvature. 

In the case of a degenerate ground-state manifold, (relevant for topologically-ordered ground states), one deals with a ground-state multiplet~\cite{hatsugai2005characterization} 
\begin{align}
\bs{\Psi}_{{\rm GS}} = \left ( \vert \Psi_0 \rangle, \vert \Psi_1 \rangle , \dots  \vert \Psi_q \rangle \right ).
\end{align}
The components of the non-Abelian Berry connection (a $q \times q$ matrix) are then given by
\begin{align}
&\left (\mathcal{A}_{\mu} \right )_{ij} = i \langle \Psi_i \vert \partial_{\lambda^{\mu}} \Psi_j  \rangle.\label{Berry_connection_def}
\end{align}
It is convenient to introduce the averaged Berry connection over the ground-state manifold,
\begin{align}
\mathcal{A}_{\mu} = (1/q) \, \sum_{i = 1}^q \left (\mathcal{A}_{\mu} \right )_{ii} , \label{av_Berry}
\end{align}
from which one builds the Berry curvature of the ground-state multiplet
\begin{align}
\mathcal{F}_{1 2} = \partial_{\lambda_{1}} \mathcal A_{2} - \partial_{\lambda_{2}} \mathcal A_{1}.
\end{align}
The corresponding many-body Chern number is defined by
\begin{align}
&\nu_{\rm{Ch}}^{\rm{MB}} = \frac{1}{2 \pi} \int_{\mathcal{M}} \mathcal{F}_{12} (\bs{\lambda}) \, \rm{d}^2\lambda \, \in \mathbb{Q}.  \label{MB_Chern_def}
\end{align}
This topological invariant can be a fractional number when $q\!>\!1$, due to the $(1/q)$ factor in Eq.~\eqref{av_Berry}.

The goal of the present work is to relate the Hall conductivity of insulating states of matter to the many-body Chern number $\nu_{\rm{Ch}}^{\rm{MB}}$, as defined in the parameter space of twist-boundary conditions, without resorting to the Kubo formula.

\section{Many-body Chern number defined in the space of twisted boundary phases}
\label{sec:mbcntwist}

We now consider a generic $N$-body Hamiltonian in 2D space (size $L_x \times L_y$)
\begin{align}
	\hat{H}
	=
	\sum_{a=1}^N
	\left(
	\frac{[\hat{\mathbf{p}}_a - \mathbf{A}(\hat{\mathbf{r}}_a)]^2}{2m} + V(\hat{\mathbf{r}}_a)
	\right)
	+
	\hat{H}_\mathrm{int} (\{\hat{\mathbf{r}}_a\}),\label{Ham_def} 
\end{align}
where $\hat{\mathbf{r}}_a$ and $\hat{\mathbf{p}}_a$ are the position and momentum operators for the $a$-th particle with mass $m$. The Hamiltonian also contains the magnetic vector potential $\mathbf{A}(\hat{\mathbf{r}}_a)$, the one-body potential $V(\hat{\mathbf{r}}_a)$ and the inter-particle interaction $\hat{H}_\mathrm{int}(\{\hat{\mathbf{r}}_a\})$.

For convenience, and without loss of generality, we now take the position representation so that the position operator $\hat{\mathbf{r}}_a$ is given by the position of the $a$-th particle $\mathbf{r}_a$, and the momentum operator is $\hat{\mathbf{p}}_a = -i\hbar \boldsymbol\nabla_a$. 
The wavefunction $\psi (\{ \mathbf{r}_a\})$ then depends on coordinates of $N$ particles. This wavefunction $\psi (\{ \mathbf{r}_a\})$ can generally be a spinor with multiple internal degrees of freedom, such as spins or different layers for multi-layer systems.

We now introduce generalized (twisted) boundary conditions~\cite{niu1985quantized,avron1985quantization,tao1986impurity,souza2000polarization}.
To begin with, we clarify the boundary conditions without a twist. In the presence of non-zero magnetic field, the magnetic vector potential $\mathbf{A}(\mathbf{r})$ cannot be made periodic.
Instead, the magnetic vector potential can obey the boundary conditions~\cite{FradkinBook}
\begin{align}
	A_x (x, y + L_y) &= A_x (x,y) + \hbar \partial_y \phi_x (x,y)
	\notag \\
	A_y (x + L_x, y) &= A_y (x,y) + \hbar \partial_x \phi_y (x,y), \label{eq:periodicA}
\end{align}
where the scalar functions $\phi_x (\mathbf{r})$ and $\phi_y (\mathbf{r})$ are known as transition functions. 
The lengths of the system in the $x$ and $y$ directions are denoted by $L_x$ and $L_y$, respectively. For certain lattice models, these transition functions can be chosen trivial, even in the presence of a net magnetic flux, but this is not possible in general.
In terms of these transition functions, the periodic boundary conditions correspond to imposing the following:
\begin{align}
	\hat T_{a^\prime} (L_x \bs{1}_x) \psi (\{ \mathbf{r}_a\}) &= e^{i \phi_x (\mathbf{r}_{a^\prime})} \psi (\{ \mathbf{r}_a\}) , \notag \\
	\hat T_{a^\prime} (L_y \bs{1}_y) \psi (\{ \mathbf{r}_a\}) &= e^{i \phi_y (\mathbf{r}_{a^\prime})} \psi (\{ \mathbf{r}_a\}), \quad \forall a^\prime,
	\label{eq:periodicbc}
\end{align}
where the translation operator $\hat T_{a^\prime} (\mathbf{r}_0)$ translates the position of the $a^\prime$-th particle by the amount $\mathbf{r}_0$. Here, $\bs{1}_x$ and $\bs{1}_y$ are the unit vectors in the $x$-direction and $y$-direction, respectively.
The condition in Eq.~(\ref{eq:periodicbc}) describes the fact that the wavefunction should come back to its original value as one goes around the system, up to the factor due to the non-periodic nature of the magnetic vector potential, Eq.~(\ref{eq:periodicA}).

Twisted boundary conditions are a generalization of Eq.~\eqref{eq:periodicbc}, which correspond to setting~\cite{niu1985quantized,avron1985quantization,tao1986impurity}
\begin{align}
\hat T_{a^\prime} (L_x \bs{1}_x) \psi (\{ \mathbf{r}_a\}) &= e^{i (\phi_x (\mathbf{r}_{a^\prime}) + \theta_x)} \psi (\{ \mathbf{r}_a\}), \notag \\
\hat T_{a^\prime} (L_y \bs{1}_y) \psi (\{ \mathbf{r}_a\}) &= e^{i (\phi_y (\mathbf{r}_{a^\prime}) + \theta_y)} \psi (\{ \mathbf{r}_a\}),
\end{align}
where $\theta_x$ and $\theta_y$ are the twist boundary phases that we additionally impose.

We now aim to define the many-body Chern number in the parameter space of twist phases, $(\theta_x, \theta_y) \in \mathbb{T}^2$. However, at this stage, these phases appear in the boundary conditions and not in the Hamiltonian [see Section~\ref{sec:mbcn}]. 
We can instead move the phases $(\theta_x, \theta_y)$ from the boundary conditions to the Hamiltonian by a gauge transformation.
Defining the gauge-transformed wavefunction by
\begin{align}
	\psi^\prime (\{ \mathbf{r}_a\}) \equiv
	e^{-i\theta_x\sum_{a} x_a/L_x - i\theta_y\sum_{a} y_a/L_y}\psi(\{ \mathbf{r}_a\}),
\end{align}
one obtains the gauge-transformed Hamiltonian:
\begin{align}
	&\hat{H}^\prime \equiv \\
	&e^{-i\theta_x\sum_{a} x_a/L_x - i\theta_y\sum_{a} y_a/L_y}
	\hat{H}
	e^{i\theta_x\sum_{a} x_a/L_x + i\theta_y\sum_{a} y_a/L_y}. \notag
\end{align}
The Hamiltonian in this new gauge, $\hat{H}^\prime$, is nothing but the original Hamiltonian $\hat{H}$ in Eq.~\eqref{Ham_def} with the magnetic vector potential replaced by
\begin{equation}
A_x \longrightarrow A_x - \theta_x/L_x , \quad A_y \longrightarrow A_y - \theta_y/L_y . \label{eq:gaugeA}
\end{equation}
The boundary conditions for the gauge-transformed wavefunction $\psi^\prime (\{ \mathbf{r}_a\})$ then reduces to the original periodic boundary condition in Eq.~(\ref{eq:periodicbc}).
Thus, the gauge transformation brings us back to a standard torus geometry, with usual periodic boundary conditions, but now introduces $\theta_{x,y}$ as fluxes going through the torus~\cite{niu1985quantized,avron1985quantization,tao1986impurity}; see Eq.~\eqref{eq:gaugeA} and Fig.~\ref{fig:illustration2}.
We hereby denote $\hat{H}(\theta_x, \theta_y)$, the Hamiltonian describing our original system in the presence of these additional fluxes, i.e.~the original Hamiltonian $\hat{H}$ in Eq.~(\ref{Ham_def}) with the magnetic vector potential replaced by $\theta_{x,y}$-dependent factors as in Eq.~(\ref{eq:gaugeA}).

Now that the Hamiltonian is parametrized as $\hat H (\theta_x, \theta_y)$, the relevant parameter space is $\boldsymbol\theta \equiv (\theta_x,\theta_y)\in \mathbb{T}^2$, and the gapped many-body ground states $\Psi_n (\{ \mathbf{r}_a\}, \boldsymbol\theta)$, with possible $q$-fold degeneracy, are defined over the two-dimensional ``flux" space. In terms of these many-body ground states, we define the many-body Chern number [Eq.~\eqref{MB_Chern_def}]
\begin{align}
&\nu_{\rm{Ch}}^{\rm{MB}} = \frac{1}{2 \pi} \int_{\mathbb{T}^2} \mathcal{F}_{xy} (\bs{\theta}) \, \rm{d}^2\theta \in \mathbb{Q} ,  \notag
\end{align}
where
\begin{align}
\mathcal{F}_{xy} (\bs{\theta}) &= \partial_{\theta_{x}} \mathcal A_{y} (\boldsymbol\theta) - \partial_{\theta_{y}} \mathcal A_{x} (\boldsymbol\theta), \\
\mathcal{A}_{\mu}(\boldsymbol\theta) &= \frac{i}{q}\sum_{n = 1}^q \langle \Psi_n (\boldsymbol\theta) \vert \partial_{\theta_{\mu}} \Psi_n (\boldsymbol\theta) \rangle,
\end{align}
are, respectively, the many-body Berry curvature and the Berry connection defined over the twist-angle space; see Eqs.~\eqref{Berry_connection_def}-\eqref{MB_Chern_def}. Here we introduced back the ``ket'' notation $|\Psi_n (\boldsymbol\theta)\rangle$, whose position-basis representation is the many-body wavefunction $\langle \{ \mathbf{r}_a\} | \Psi_n (\boldsymbol\theta)\rangle = \Psi_n (\{ \mathbf{r}_a\},\boldsymbol\theta)$.

\section{Relating the many-body Chern number to a circular-dichroic response:~Using Fermi's Golden rule}
\label{sec:cd}

\begin{figure}[h!]
{\scalebox{0.34}
{\includegraphics{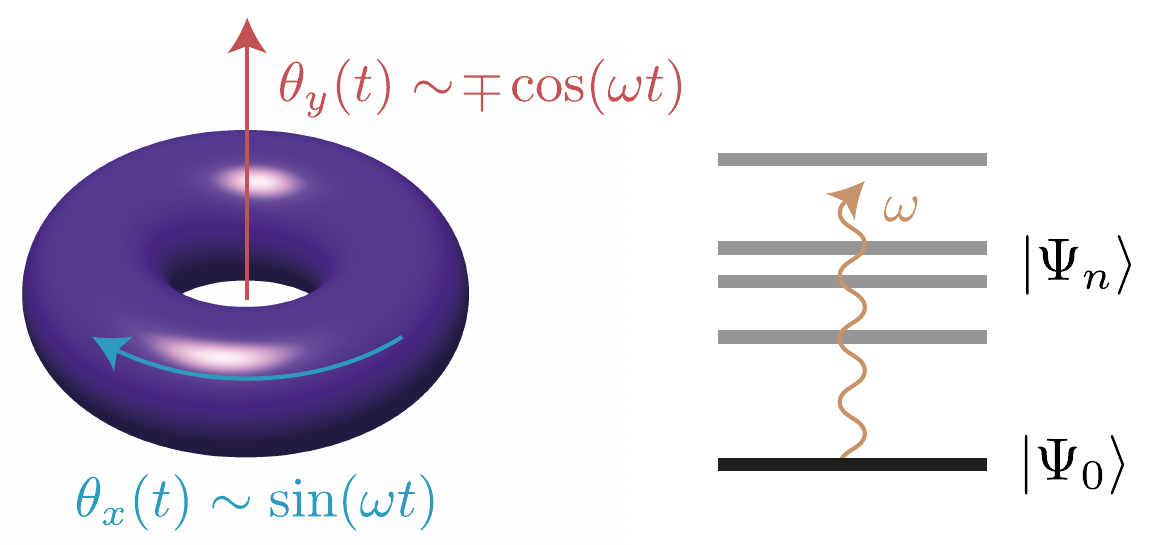}}}
\caption{Circular dichroism as induced by time-dependent fluxes through the torus: The excitation rate (and hence, the absorbed energy) depends on the sign ($\mp$) of the perturbing fields.}
\label{fig:illustration2}
\end{figure}

We now relate the many-body Chern number to a relevant physical observable:~excitation rates upon a circular drive in twist-angle (or flux) space. The Hamiltonian in Eq.~\eqref{Ham_def} is now defined on a regular torus, with periodic boundary conditions [Eq.~\eqref{eq:periodicbc}], and we hereby set $L_{x,y}\!=\!L$. Let us now prepare the system in the gapped ground state $\vert \Psi_0 \rangle$, in the absence of flux $\theta_{x,y}$. (If the ground state is degenerate, we consider one of the ground states at a time.) Let us study its response to a circular electric field:
\begin{equation}
\bs{E}_{\pm} (\omega) = 2 \mathcal{E} \left [ \cos (\omega t) , \pm \sin (\omega t) \right ] = - \partial_t \bs{A}. \label{electric_field}
\end{equation}
The upper (lower) sign corresponds to the right (left) circularly polarized field. 
In terms of the magnetic vector potential, this perturbing driving field corresponds to adding the following terms to its components:
\begin{equation}
A_x \longrightarrow A_x - (2\mathcal{E}/\omega) \sin (\omega t) , \quad A_y \longrightarrow A_y \pm (2\mathcal{E}/\omega) \cos (\omega t) .\notag
\end{equation}
It is insightful to interpret these additional terms in the magnetic vector potential using the twist-angle picture [Eq.~\eqref{eq:gaugeA}]:~this corresponds to preparing the system in the state $\vert \Psi_0 (\bs{\theta}=0)\rangle$ and to evolve this state according to the time-dependent Hamiltonian $\hat H (\theta_x(t),\theta_y(t)) $, where
\begin{align}
&\theta_x(t) = (2\mathcal{E} L/\omega) \sin (\omega t) , \\
&\theta_y(t) = \mp (2\mathcal{E} L/\omega) \cos (\omega t), \notag 
\end{align}
are interpreted as time-dependent fluxes; see Fig.~\ref{fig:illustration2} for an illustration of this setting.

Assuming a weak field such that the final excited fraction is small, we calculate the excited fraction using the Fermi's golden rule, i.e.~using lowest-order time-dependent perturbation theory~\cite{GriffithsBook}. Here, the perturbation takes the form
\begin{align}
	\delta \hat{H} =& \left.\hat{H}(\theta_x, \theta_y)\right|_{(\theta_x, \theta_y) = ((2\mathcal{E} L/\omega) \sin (\omega t), \mp (2\mathcal{E} L/\omega) \cos (\omega t))} \notag \\ &- \hat{H}(0, 0)
	\notag \\
	=&
	\frac{2\mathcal{E}L}{\omega}
	\left[
	\left.\frac{\partial \hat{H}(\boldsymbol\theta)}{\partial \theta_x}\right|_{\boldsymbol\theta = 0}\sin (\omega t)
	\mp
	\left. \frac{\partial \hat{H}(\boldsymbol\theta)}{\partial \theta_y}\right|_{\boldsymbol\theta = 0}\cos (\omega t)
	\right]
	\notag \\
	&
	+\mathcal{O}(\mathcal{E}^2)
	\notag \\
	=&
	\frac{\mathcal{E}L}{\omega}
	\left.
	\left(
	\frac{1}{i}\frac{\partial \hat{H}(\boldsymbol\theta)}{\partial \theta_x}
	\mp
	\frac{\partial \hat{H}(\boldsymbol\theta)}{\partial \theta_y}
	\right)
	\right|_{\boldsymbol\theta = 0}
	e^{i\omega t} + c.c.
	\notag \\
	&
	+\mathcal{O}(\mathcal{E}^2),
\end{align}
where the Taylor expansion is allowed due to the weakness of the perturbation. Then, the transition probability at time $t$ from the original state to any of the excited states can be calculated from Fermi's golden rule~\cite{GriffithsBook},
\begin{align}
n_{\rm{ex}}^{\pm}(t)=& \frac{2 \pi t}{\hbar} \sum_{n \ne 0}  \left|
	\frac{\mathcal E L}{\hbar \omega}
	\bigg \langle \Psi_n \bigg | \frac{1}{i}\frac{\partial \hat{H}}{\partial \theta_{x}} \mp \frac{\partial \hat{H}}{\partial \theta_{y}} \bigg |\Psi_0 \bigg \rangle
	\right|^2_{\bs{\theta}=0}
	\notag \\
	&\cdot \delta^{(t)} (E_n - E_0 - \hbar \omega),
\end{align}
where the function $\delta^{(t)}(\varepsilon) \equiv (2\hbar /\pi t)\sin^2(\varepsilon t/2\hbar)/\varepsilon^2$ approaches the delta function in the long time limit: $\delta^{(t)}(\varepsilon) \to \delta (\varepsilon)$ as $t \to \infty$. 
A priori, the sum is performed over all many-body excited states.
However, we point out that this excited fraction only involves single-particle excitations of the many-body ground state, since the operators $\partial_{\theta_i} \hat H$ are single-particle operators.

Defining the excitation rates $\Gamma^{\pm} = n_{\rm{ex}}^{\pm}(t) / t$, one finds~\cite{tran2017probing,ozawa2019probing}
\begin{align}
\Delta \Gamma^{\rm{int}}=&\frac{1}{2}\int_0^{\infty} \Gamma^{+}(\omega) - \Gamma^{-}(\omega) \, {\rm d} \omega \notag \\
=& \sum_{n \ne 0} \frac{\mathcal{E}^2 L^2 \pi}{\hbar^2 (E_n - E_0)^2} \left\{ \left|
	\left\langle \Psi_n \left| \frac{1}{i}\frac{\partial \hat{H}}{\partial \theta_{x}} - \frac{\partial \hat{H}}{\partial \theta_{y}} \right|\Psi_0 \right\rangle
	\right|^2_{\bs{\theta}=0} \right.
	\notag \\
	&-
	\left. \left|
	\left\langle \Psi_n \left| \frac{1}{i}\frac{\partial \hat{H}}{\partial \theta_{x}} + \frac{\partial \hat{H}}{\partial \theta_{y}} \right|\Psi_0 \right\rangle
	\right|^2_{\bs{\theta}=0} \right\}
	\notag \\
	=&
	\sum_{n \ne 0} \frac{\mathcal{E}^2 L^2 2\pi i}{\hbar^2 (E_n - E_0)^2}
	\left(
	\langle \Psi_0 | \frac{\partial \hat{H}}{\partial \theta_{x}}|\Psi_n \rangle
	\langle \Psi_n | \frac{\partial \hat{H}}{\partial \theta_{y}}|\Psi_0 \rangle
	\right.
	\notag \\
	&-
	\left.\left.
	\langle \Psi_0 | \frac{\partial \hat{H}}{\partial \theta_{y}}|\Psi_n \rangle
	\langle \Psi_n | \frac{\partial \hat{H}}{\partial \theta_{x}}|\Psi_0 \rangle
	\right)\right|_{\boldsymbol\theta = 0}
	\notag \\
=& \left (\frac{\mathcal{E}^2L^2}{\hbar^2} \right ) \times 2 \pi \mathcal{F}_{xy} (\bs{\theta}=0), \label{result_FGR}
\end{align}
where we used the expression for the Berry curvature in Eq.~\eqref{eq:berrycurvature2} at the final step.
Hence, we obtained that the dichroic response yields the Berry curvature at the origin of flux space (i.e.~in the absence of flux/twist).

In the thermodynamic limit, any $\bs{\theta}$-dependence (boundary effects) vanishes~\cite{kudo2019many}, hence one obtains a relation between the Berry curvature and the many-body Chern number:
\begin{align}
&2 \pi \mathcal{F}_{xy} (\bs{\theta}=0) \\
&\longrightarrow 2 \pi \langle \mathcal{F}_{xy} \rangle_{\bs{\theta}} = 2 \pi \times \frac{1}{(2\pi)^2} \int_{\mathbb{T}^2} \mathcal{F}_{xy}(\bs{\theta}) \, {\rm d}^2\theta = \nu_{\rm{Ch}}^{\rm{MB}} . \notag
\end{align}
Consequently, one obtains from Eq.~\eqref{result_FGR} that 
\begin{equation}
\Delta \Gamma^{\text{int}}/A = (\mathcal{E}/\hbar)^2 \times \nu_{\text{Ch}}^{\text{MB}},\label{Chern_rates}
\end{equation}
is a (quantized) topological response, where $A \equiv L^2$ is the area of the system.  This result illustrates how the many-body Chern number of correlated insulators can be determined by monitoring single-particle excitations upon a circular drive, within a Fermi-golden rule framework. This conceptually simple picture has practical implications for quantum-engineered systems, where excitation rates can be directly monitored~\cite{asteria2019measuring,tan2019experimental,yu2020experimental,chen2022synthetic}.

For a Chern insulator (non-interacting fermions): 
\begin{equation}
\nu_{\rm{Ch}}^{\rm{MB}} = \nu_{\rm{Ch}} = \frac{1}{2 \pi} \int_{\rm{FBZ}} \mathcal{F} (\bs{k}) \, \rm{d}^2 k \in \mathbb{Z}  ,\notag
\end{equation}
such that $\Delta \Gamma^{\text{int}}$ is quantized to an integer, where $\mathcal{F} (\mathbf{k})$ is the Berry curvature in momentum space, and $\nu_\mathrm{Ch}$ is the momentum-space Chern number of the occupied Bloch band. In Appendix, we provide a derivation of the relation $\nu_{\rm{Ch}}^{\rm{MB}}\!=\!\nu_{\rm{Ch}}$, hence connecting the momentum-space and twist-angle formulations in this non-interacting case.
The relation between the dichroic response and the Chern number for non-interacting fermions was experimentally measured in ultracold atomic gases in Hamburg~\cite{asteria2019measuring}.

For a fractional Chern insulator: 
\begin{equation}
\nu_{\rm{Ch}}^{\rm{MB}} \in \mathbb{Q} \longrightarrow \Delta \Gamma^{\text{int}} \text{\, is fractionally quantized !} \notag
\end{equation}
This was numerically demonstrated in Ref.~\cite{repellin2019detecting}.\\

\section{Relating the dichroic response to the Hall conductivity:~Using Kramers-Kronig relations}
\label{sec:kk}

Now we relate the Hall conductivity $\sigma_{{\rm H}}$ to the dichroic response, and thus to the many-body Chern number $\nu_{\rm{Ch}}^{\rm{MB}}$. 

We previously found, using Fermi's golden rule,
\begin{equation}
\Delta \Gamma^{\rm{int}}/A=\frac{1}{2A}\int_0^{\infty} \Gamma^{+}(\omega) - \Gamma^{-}(\omega) \, {\rm d} \omega = \left (\mathcal{E}/\hbar \right )^2 \times \nu_{\rm{Ch}}^{\rm{MB}} , \notag
\end{equation}

The excitation rate $\Gamma^\pm (\omega)$ under the circular electric field can be written in terms of the {\it absorption power} $P_\pm (\omega)$ as
\begin{align}
P_\pm(\omega) = \hbar \omega \Gamma^\pm (\omega). \label{eq:pab1}
\end{align}
On the other hand, the power absorbed upon the circular electric field is related to the conductivity tensor; this was shown in Ref.~\cite{bennett1965faraday}, and we briefly reproduce this result here.
The circular electric field that we apply in Eq.~\eqref{electric_field} can be written as
\begin{align}
	\mathbf{E}_\pm (\omega)
	=
	\left[ 1, \mp i \right]\mathcal{E} e^{i\omega t} + c.c.
\end{align}
The linear coefficient of the current $J_i$ in the $i$-direction in response to the electric field $E_j e^{i\omega t}$ in the $j$-direction is nothing but the conductivity $\sigma_{ij}$.
Therefore, the current induced by this electric field is
\begin{align}
	\mathbf{J}_\pm (\omega)
	=&
	\left[ \sigma_{xx}(\omega) \mp i \sigma_{xy}(\omega), \sigma_{yx}(\omega) \mp i \sigma_{yy}(\omega) \right]\mathcal{E} e^{i\omega t} \notag \\
	&+ c.c.
\end{align}
The absorbed power per volume caused by this electric field is then given by
\begin{align}
	&P_\pm(\omega)/A
	=
	\mathbf{J}_\pm (\omega) \cdot \mathbf{E}_\pm (\omega)
	\notag \\
	&\approx
	\left\{ \left( \sigma_{xx}(\omega) \mp i \sigma_{xy}(\omega)\right)
	\pm i
	\left( \sigma_{yx}(\omega) \mp i \sigma_{yy}(\omega)\right) \right\} \mathcal{E}^2
	+ c.c.
	\notag \\
	&= \left\{ 2\, \mathrm{Re} \left[\sigma_{xx}(\omega) + \sigma_{yy}(\omega)\right] \pm 4\, \mathrm{Im}\left[\sigma_{xy}(\omega)\right] \right\} \mathcal{E}^2,
	\label{eq:pab2}
\end{align}
where at the second line we ignored terms that oscillate as $e^{\pm 2i\omega t}$ because these terms average to zero over long time. 
At the third line, we also used $\sigma_{yx}(\omega) = -\sigma_{xy}(\omega)$, which holds when the material possesses the rotational symmetry; we note that the result readily generalizes to non-isotropic situations.
Combining Eq.~\eqref{eq:pab1} and Eq.~\eqref{eq:pab2}, we obtain
\begin{align}
P_{\pm}(\omega)&= \hbar \omega \times \Gamma^{\pm} (\omega) \\
&= 4 A \mathcal{E}^2 \left\{ \mathrm{Re} \left[\sigma_{xx}(\omega) + \sigma_{yy}(\omega)\right]/2 \pm \mathrm{Im}\left[\sigma_{xy}(\omega)\right] \right\}. \notag
\end{align}

Hence, the dissipative optical conductivity, $\mathrm{Im}[\sigma_{ij}(\omega)]$, relates to the difference in the excitation rate $\Delta \Gamma (\omega) \equiv \Gamma^+ (\omega) - \Gamma^- (\omega)$ according to 
\begin{equation}
\mathrm{Im}[\sigma_{xy} (\omega)] = \hbar \omega \Delta \Gamma (\omega) / 8 A \mathcal{E}^2 . \label{eq:imsigmaxy}
\end{equation}

Besides, we have the sum rule provided by Kramers-Kronig relations~\cite{kronig1926theory,kramers1927diffusion,bennett1965faraday,hu1989kramers,jackson1999classical,souza2008dichroic,tran2017probing}
\begin{equation}
\sigma_{\rm{H}} =\lim_{\omega\rightarrow0}\mathrm{Re}\left[ \sigma_{x y} (\omega) \right]=\frac{2}{\pi}  \int_{0}^{\infty} \frac{\mathrm{Im}\left[ \sigma_{x y} (\tilde \omega)\right]}{\tilde{\omega}} \, \rm{d} \tilde \omega , \label{eq:kk}
\end{equation}
where $\sigma_H$ is the Hall conductivity. 
Then, dividing both sides of Eq.~\eqref{eq:imsigmaxy} by $\omega$ and integrating over $\omega$, and using the sum rule in Eq.~\eqref{eq:kk}, we obtain a relation between the absorption rates and the Hall conductivity
\begin{equation}
 \Delta \Gamma^{\rm{int}}/A=\frac{1}{2A}\int_0^{\infty} \Delta \Gamma(\omega) \, {\rm d} \omega = (2\pi \mathcal{E}^2/\hbar) \, \sigma_{\rm{H}} . \notag
\end{equation}

Finally, using the main result of the previous Section, Eq.~\eqref{Chern_rates}, one obtains ~\cite{tran2017probing,repellin2019detecting}
\begin{equation}
\sigma_{{\rm H}}/\sigma_0 = \nu_{\rm{Ch}}^{\rm{MB}} , \label{final_equation}
\end{equation}
where  $\sigma_0\!=\!(1/h)$ is the quantum of conductance $(e=1)$. Equation~\eqref{final_equation} is the quantization of the Hall conductivity according to the many-body Chern number, valid for any 2D gapped system. 

\section{Some concluding remarks}
\label{sec:conclusion}

Here we summarize the take-home messages of this pedagogical piece.\\ 

Take-home message 1: The dichroic response offers a probe for the Hall response, which does not rely on any transport experiment~\cite{tran2017probing,schuler2020local,schuler2020circular}
\begin{equation}
 \Delta \Gamma^{\rm{int}}/A= (2\pi \mathcal{E}^2/\hbar) \, \sigma_{\rm{H}}, \notag
\end{equation}
as explored in the ultracold gas experiment in Hamburg~\cite{asteria2019measuring}.
This relation is essentially a consequence of Kramers-Kronig relations~\cite{kronig1926theory,kramers1927diffusion,bennett1965faraday,hu1989kramers,jackson1999classical,souza2008dichroic,tran2017probing}.
We note that another alternative to transport is provided by the Streda-formula approach:~a density response to an external magnetic field also provides the quantized Hall response of correlated insulators~\cite{repellin2020fractional,leonard2023realization,gavensky2023connecting}.

Take-home message 2: The dichroic response reflects the many-body Chern number
\begin{equation}
\Delta \Gamma^{\rm{int}}/A= \left (\mathcal{E}/\hbar \right )^2 \times \nu_{\rm{Ch}}^{\rm{MB}} .\notag
\end{equation}
This result follows from Fermi's golden rule.
In particular, circular dichroism offers a topological signature for FQH states~\cite{repellin2019detecting}, where $\nu_{\rm{Ch}}^{\rm{MB}}$ becomes fractional, which is particularly appealing for recent cold-atom realizations~\cite{leonard2023realization}.

Take-home message 3: Combining the above two relations, we obtain the relation connecting the quantized Hall conductivity to the many-body Chern number
\begin{equation}
\sigma_{{\rm H}}/\sigma_0 = \nu_{\rm{Ch}}^{\rm{MB}} \in \mathbb{Q},  \text{  (in gapped 2D system),} \notag
\end{equation}
previously derived using the Kubo formula~\cite{niu1985quantized,tao1986impurity,avron1985quantization}.

We believe that the derivation and the viewpoint presented in this pedagogical piece offer a fresh interpretation to the well-established relation between the Hall conductivity and the many-body Chern number, and that they will further motivate the search for fundamental relations between topological invariants and physical response functions.

\begin{acknowledgements}
We  acknowledge various stimulating exchanges with our collaborators over the years. Special thanks go to Luca Asteria, Jianming Cai, Nigel Cooper, Jean Dalibard, Alexandre Dauphin, Lucila Peralta Gavensky, Adolfo G.~Grushin, Bruno Mera, Cecile Repellin, Duc Thanh Tran, Christof Weitenberg and Peter Zoller. NG is supported by the ERC Grant LATIS and the EOS project CHEQS. TO is supported by JSPS KAKENHI Grant No. JP20H01845, JST PRESTO Grant No. JPMJPR2353, and JST CREST Grant No. JPMJCR19T1.
\end{acknowledgements}

\appendix*

\section{Relation between the many-body Chern number and the momentum-space Chern number for noninteracting fermions}

Here we provide a brief proof that the many-body Chern number $\nu_\mathrm{Ch}^\mathrm{MB}$ reduces to the conventional momentum-space Chern number~\cite{thouless1982quantized} -- the integral of the momentum-space Berry curvature over the first Brillouin zone -- in the case of a non-interacting Chern insulator,~i.e.~a filled band of non-interacting fermions.

Let us first define the momentum-space Chern number. We consider the $N$-body Hamiltonian introduced in Eq.~(\ref{Ham_def}) without the interaction term, namely
\begin{align}
	\hat{H}
	=
	\sum_{a=1}^N
	\left(
	\frac{[\hat{\mathbf{p}}_a - \mathbf{A}(\hat{\mathbf{r}}_a)]^2}{2m} + V(\hat{\mathbf{r}}_a)
	\right).
\end{align}
We denote the corresponding single-particle Hamiltonian as
\begin{align}
	\hat{H}^{(1)}
	=
	\frac{[\hat{\mathbf{p}} - \mathbf{A}(\hat{\mathbf{r}})]^2}{2m} + V(\hat{\mathbf{r}}).
\end{align}
Assuming a periodic lattice structure, the Bloch theorem indicates that the eigenvalues and eigenstates are labeled by the band index $n$ and the quasimomentum $\mathbf{k} = (k_x, k_y)$:
\begin{align}
	\hat{H}^{(1)}|\chi_{n,\mathbf{k}}\rangle = \varepsilon_n (\mathbf{k}) |\chi_{n,\mathbf{k}}\rangle,
\end{align}
where the Bloch wavefunction $|\chi_{n,\mathbf{k}}\rangle$ can be written, in position representation, as $\chi_{n,\mathbf{k}}(\mathbf{r}) \equiv \langle \mathbf{r}|\chi_{n,\mathbf{k}}\rangle = e^{i\mathbf{k}\cdot\mathbf{r}}u_{n,\mathbf{k}}(\mathbf{r})$ with $u_{n,\mathbf{k}}(\mathbf{r})$ obeying the  periodicity of the original lattice (or a generalized `magnetic' periodicity if a nonzero magnetic flux is present).

We consider that the first band $n\!=\!1$ is fully filled by non-interacting fermions. In this case, the many-body Chern number $\nu_\mathrm{Ch}^\mathrm{MB}$ reduces to the momentum-space Chern number introduced by TKNN~\cite{thouless1982quantized}, as we now explain.

The topological properties of the first Bloch band are defined with respect to the cell-periodic part of the Bloch wavefunction, $u_{1,\mathbf{k}}(\mathbf{r})$.
We note that the cell-periodic part of the Bloch wavefunction is an eigenstate of the $\mathbf{k}$-space Hamiltonian defined by
\begin{align}
	&\hat{H}(\mathbf{k}) \equiv e^{-i\mathbf{k}\cdot \hat{\mathbf{r}}} \hat{H}^{(1)} e^{i\mathbf{k}\cdot \hat{\mathbf{r}}}
	=
	\frac{[\hat{\mathbf{p}} - \mathbf{A}(\hat{\mathbf{r}}) + \mathbf{k}]^2}{2m} + V(\hat{\mathbf{r}});
	\\
	&\hat{H}(\mathbf{k})|u_{n,\mathbf{k}}\rangle = \varepsilon_n (\mathbf{k}) |u_{n,\mathbf{k}}\rangle,
\end{align}
where we have introduced the ``ket" notation, $u_{n,\mathbf{k}}(\mathbf{r}) = \langle \mathbf{r}|u_{n}(\mathbf{k})\rangle$ for the cell-periodic part of the Bloch wavefunction.

Following Eqs.~\eqref{eq:berrycurvature} and~\eqref{eq:berrycurvature2}, the Berry curvature in the first band, defined over $\mathbf{k}$-space, can be written as
\begin{align}
	\mathcal{F}_{12}& (\mathbf{k})
	= i  \left( \, \left\langle \frac{\partial u_{1, \mathbf{k}}}{\partial k_x} \right| \left. \frac{\partial u_{1, \mathbf{k}}}{\partial k_y} \right\rangle - \left\langle \frac{\partial u_{1, \mathbf{k}}}{\partial k_y} \right| \left. \frac{\partial u_{1, \mathbf{k}}}{\partial k_x} \right\rangle  \right)
	\notag \\
	=& i \sum_{n \neq 1} \frac{1}{(\varepsilon_n (\mathbf{k}) - \varepsilon_1 (\mathbf{k}))^2} \cdot 
	\notag \\
	&	
	\left( \, \langle u_{1, \mathbf{k}} | \frac{\partial \hat{H}(\mathbf{k})}{\partial k_x} | u_{n, \mathbf{k}}\rangle\langle u_{n, \mathbf{k}} | \frac{\partial \hat{H}(\mathbf{k})}{\partial k_y}| u_{1, \mathbf{k}}\rangle
	\right.
	\notag \\
	&-
	\left.
	\langle u_{1, \mathbf{k}} | \frac{\partial \hat{H}(\mathbf{k})}{\partial k_y} | u_{n, \mathbf{k}} \rangle\langle u_{n, \mathbf{k}} | \frac{\partial \hat{H}(\mathbf{k})}{\partial k_x}| u_{1, \mathbf{k}} \rangle  \right).
\end{align}
The momentum-space Chern number $\nu_\mathrm{Ch}$ is then defined in terms of $\mathcal{F}_{12}(\mathbf{k})$ as 
\begin{equation}
\nu_\mathrm{Ch} = \frac{1}{2\pi}\int d^2 k \, \mathcal{F}_{12}(\mathbf{k}).
\end{equation}

We will now relate the many-body Berry curvature $\mathcal{F}^\mathrm{MB}_{12}$ defined over the space of twisted boundary conditions to the $\mathbf{k}$-space Berry curvature $\mathcal{F}_{12} (\mathbf{k})$. The many-body Berry curvature is twist-angle space reads
\begin{align}
	\mathcal{F}^\mathrm{MB}_{12}&(\boldsymbol\theta)
	=
	i\sum_{N \neq 0} \frac{1}{(E_N - E_0)^2}\cdot
	\notag \\
	&
	\left(
	\langle \Psi_0 | \frac{\partial \hat{H}^\mathrm{MB}(\boldsymbol\theta)}{\partial \theta_{x}}|\Psi_N \rangle
	\langle \Psi_N | \frac{\partial \hat{H}^\mathrm{MB}(\boldsymbol\theta)}{\partial \theta_{y}}|\Psi_0 \rangle
	\right.
	\notag \\
	&-
	\left.
	\langle \Psi_0 | \frac{\partial \hat{H}^\mathrm{MB}(\boldsymbol\theta)}{\partial \theta_{y}}|\Psi_N \rangle
	\langle \Psi_N | \frac{\partial \hat{H}^\mathrm{MB}(\boldsymbol\theta)}{\partial \theta_{x}}|\Psi_0 \rangle
	\right),\label{MB_curvature_appendix}
\end{align}
where $|\Psi_0\rangle$ is the many-body groundstate with energy $E_0$, corresponding to the fully occupied first Bloch band; $|\Psi_N\rangle$ is any many-body eigenstate of $\hat{H}(\boldsymbol\theta)$ with energy $E_N$, which is different from $|\Psi_0\rangle$.
In this Appendix, we have added a superscript $^\mathrm{MB}$ in the many-body Berry curvature $\mathcal{F}^\mathrm{MB}_{12}(\boldsymbol\theta)$ and the many-body Hamiltonian $\hat{H}^\mathrm{MB}(\boldsymbol\theta)$ to distinguish them from single particle ones.
The many-body Hamiltonian $\hat{H}^\mathrm{MB}(\boldsymbol\theta)$ is defined by
\begin{align}
	\hat{H}^\mathrm{MB}(\boldsymbol\theta)
	=&
	\sum_{a=1}^N
	\left(
	\frac{[\hat{\mathbf{p}}_a - \mathbf{A}(\hat{\mathbf{r}}_a) + \boldsymbol\theta / L]^2}{2m} + V(\hat{\mathbf{r}}_a)
	\right)
	\notag \\
	&+
	\hat{H}_\mathrm{int} (\{\mathbf{r}_a\}),
\end{align}

As previously noted, we have
\begin{align}
	\frac{\partial \hat{H}^\mathrm{MB}(\boldsymbol\theta)}{\partial \theta_{i}}
	=
	\frac{1}{L}
	\sum_{a=1}^N
	\frac{\hat{p}_{a,i} - A_i (\hat{\mathbf{r}}_a) + \theta_i / L}{m},
\end{align}
which only involves single-particle operators; hence, matrix elements such as $\langle \Psi_N | \partial_{\theta_i} \hat{H}^{\mathrm{MB}}(\boldsymbol\theta) | \Psi_0\rangle$ are only nonzero for states $|\Psi_N\rangle$ obtained by removing one particle from the occupied band in $|\Psi_0\rangle$ and placing one particle somewhere in the upper bands. 

Let us consider a state $|\Psi_N\rangle$, which is obtained by removing a particle from the state $|\chi_{1,\mathbf{k}_0}\rangle$ from the occupied band and adding a particle into the state $|\chi_{n,\mathbf{k}_1}\rangle$.
Then,
\begin{align}
	&\langle \Psi_N | \frac{\partial \hat{H}^\mathrm{MB}(\boldsymbol\theta)}{\partial \theta_{i}}|\Psi_0 \rangle
	=
	\frac{1}{L}\langle \chi_{n,\mathbf{k}_1}|
	\frac{\hat{p}_{i} - A_i (\hat{\mathbf{r}}) + \theta_i / L}{m}
	|\chi_{1,\mathbf{k}_0}\rangle
	\notag \\
	&=
	\frac{\delta_{\mathbf{k}_0,\mathbf{k}_1}}{L}\langle u_{n,\mathbf{k}_0}|
	\frac{\hat{p}_{i} - A_i (\hat{\mathbf{r}}) + \theta_i / L + k_{0,i}}{m}
	|u_{1,\mathbf{k}_0}\rangle
	\notag \\
	&=
	\frac{\delta_{\mathbf{k}_0,\mathbf{k}_1}}{L}\langle u_{n,\mathbf{k}_0}|
	\left.\frac{\partial \hat{H}(\mathbf{k})}{\partial k_i}\right|_{\mathbf{k} = \mathbf{k}_0 + \boldsymbol\theta/L}
	|u_{1,\mathbf{k}_0}\rangle.
\end{align}
Then, since $E_N - E_0 = \varepsilon_n(\mathbf{k}_1) - \varepsilon_1 (\mathbf{k}_0)$, the many-body Berry curvature at $\boldsymbol\theta = (0,0)$ is given by [Eq.~\eqref{MB_curvature_appendix}]
\begin{align}
	\mathcal{F}^\mathrm{MB}_{12}(\boldsymbol\theta &= 0)
	=
	\frac{1}{L^2}i\sum_{n \neq 1}\sum_{\mathbf{k}} \frac{1}{(\varepsilon_n(\mathbf{k}) - \varepsilon_1 (\mathbf{k}))^2}\cdot
	\notag \\
	&
	\left(
	\langle u_{1,\mathbf{k}} | \frac{\partial \hat{H}(\mathbf{k})}{\partial k_x}|u_{n,\mathbf{k}} \rangle
	\langle u_{n,\mathbf{k}} | \frac{\partial \hat{H}(\mathbf{k})}{\partial k_y}|u_{1,\mathbf{k}} \rangle
	\right.
	\notag \\
	&-
	\left.
	\langle u_{1,\mathbf{k}} | \frac{\partial \hat{H}(\mathbf{k})}{\partial k_y}|u_{n,\mathbf{k}} \rangle
	\langle u_{n,\mathbf{k}} | \frac{\partial \hat{H}(\mathbf{k})}{\partial k_x}|u_{1,\mathbf{k}} \rangle
	\right)
	\notag \\
	&=
	\frac{1}{L^2}\sum_\mathbf{k}\mathcal{F}_{12}(\mathbf{k})
	=
	\frac{1}{(2\pi)^2}\int d^2 k\, \mathcal{F}_{12}(\mathbf{k})
	\notag \\
	&=
	\frac{1}{2\pi}\nu_\mathrm{Ch}.
\end{align}
In the thermodynamic limit, the many-body Berry curvature converges to the Chern number~\cite{kudo2019many}, $2\pi \mathcal{F}^\mathrm{MB}_{12}(\boldsymbol\theta = 0) = \nu_{\mathrm{Ch}}^\mathrm{MB}$. Hence, this establishes that $\nu_{\mathrm{Ch}}^\mathrm{MB} = \nu_\mathrm{Ch}$ for a Chern insulator, i.e.~a filled Bloch band of non-interacting fermions.

\bibliography{mybib_NG}
\bibliographystyle{ieeetr}


\end{document}